# OPERATIONAL ASPECTS OF THE MAIN INJECTOR LARGE APERTURE QUADRUPOLE (WQB)*

W. Chou#, L. Bartelson, B. Brown, D. Capista, J. Crisp, J. DiMarco, J. Fitzgerald, H. Glass, D. Harding, D. Johnson, V. Kashikhin, I. Kourbanis, P. Prieto, W. Robotham, T. Sager, M. Tartaglia, L. Valerio, R. Webber, M. Wendt, D. Wolff, M. Yang, Fermilab, Batavia, Illinois, USA

*Abstract*

A two-year Large Aperture Quadrupole (WQB) Project was completed in the summer of 2006 at Fermilab. [1] Nine WQBs were designed, fabricated and bench-tested by the Technical Division. Seven of them were installed in the Main Injector and the other two for spares. They perform well. The aperture increase meets the design goal and the perturbation to the lattice is minimal. The machine acceptance in the injection and extraction regions is increased from 40π to 60π mm-mrad. This paper gives a brief report of the operation and performance of these magnets. Details can be found in Ref [2].

## INTRODUCTION

It is known that the injection and extraction areas with Lambertson magnets are the bottleneck of the Main Injector. The physical aperture in these areas is cut to half as shown in Figure 1. The transverse acceptance is limited to 40π mm-mrad. Significant beam losses have been observed in these areas during high intensity operation.

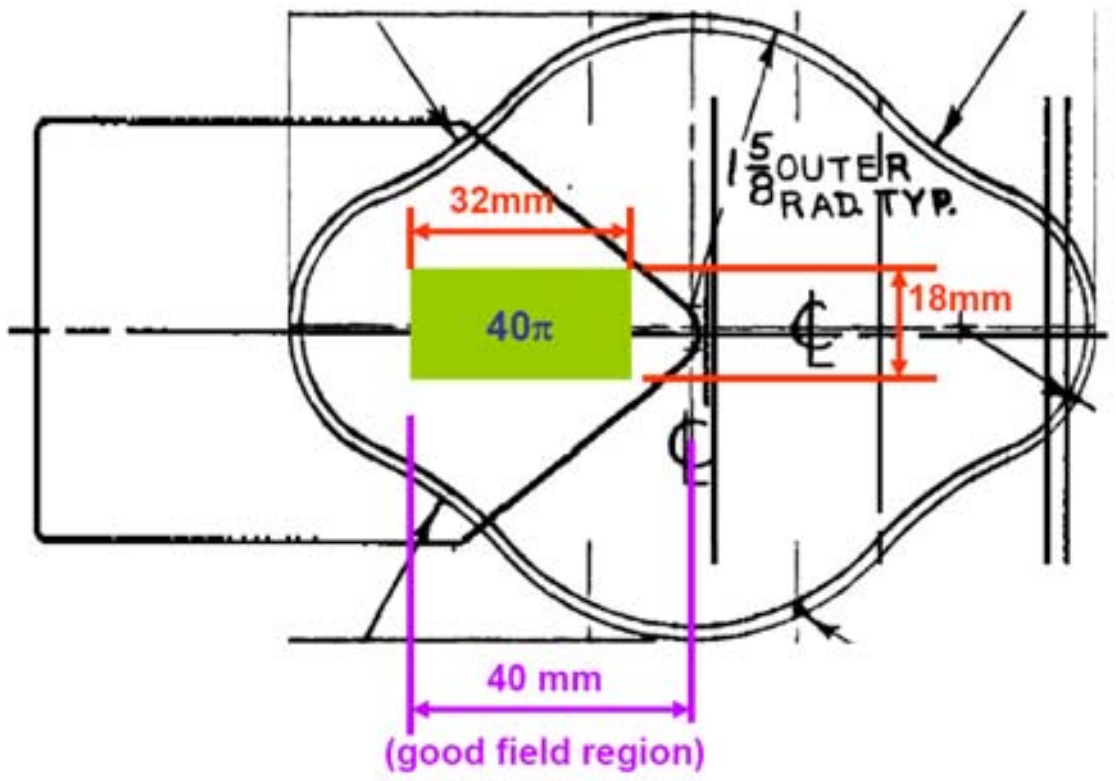

Figure 1: Physical aperture and machine acceptance in the Lambertson area.

In order to enlarge the aperture and reduce beam loss, it was decided to replace the quadrupoles in these areas by new ones (called WQB) that will have a larger aperture (4.347" vs. 3.286" of the old quad). The project started in early 2004 and was completed in the summer of 2006. Nine WQBs as well as nine new extra-wide aperture (EXWA) BPMs were designed, fabricated and tested. Seven WQBs and EXWA BPMs were installed in the Main Injector during 2006 shutdown. Their locations are listed in Table 1 and shown in Figure 2. Figure 3 shows a WQB installed in the machine.

Table 1: WQB Locations

| Location | WQB Serial No. | BPM Serial No. |
|---|---|---|
| Q101 | WQB 001 | EXWA 01 |
| Q222 | WQB 007 | EXWA 07 |
| Q321 | WQB 006 | EXWA 08 |
| Q402 | WQB 004 | EXWA 04 |
| Q522 | WQB 003 | EXWA 02 |
| Q608 | WQB 005 | EXWA 05 |
| Q620 | WQB 002 | EXWA 06 |

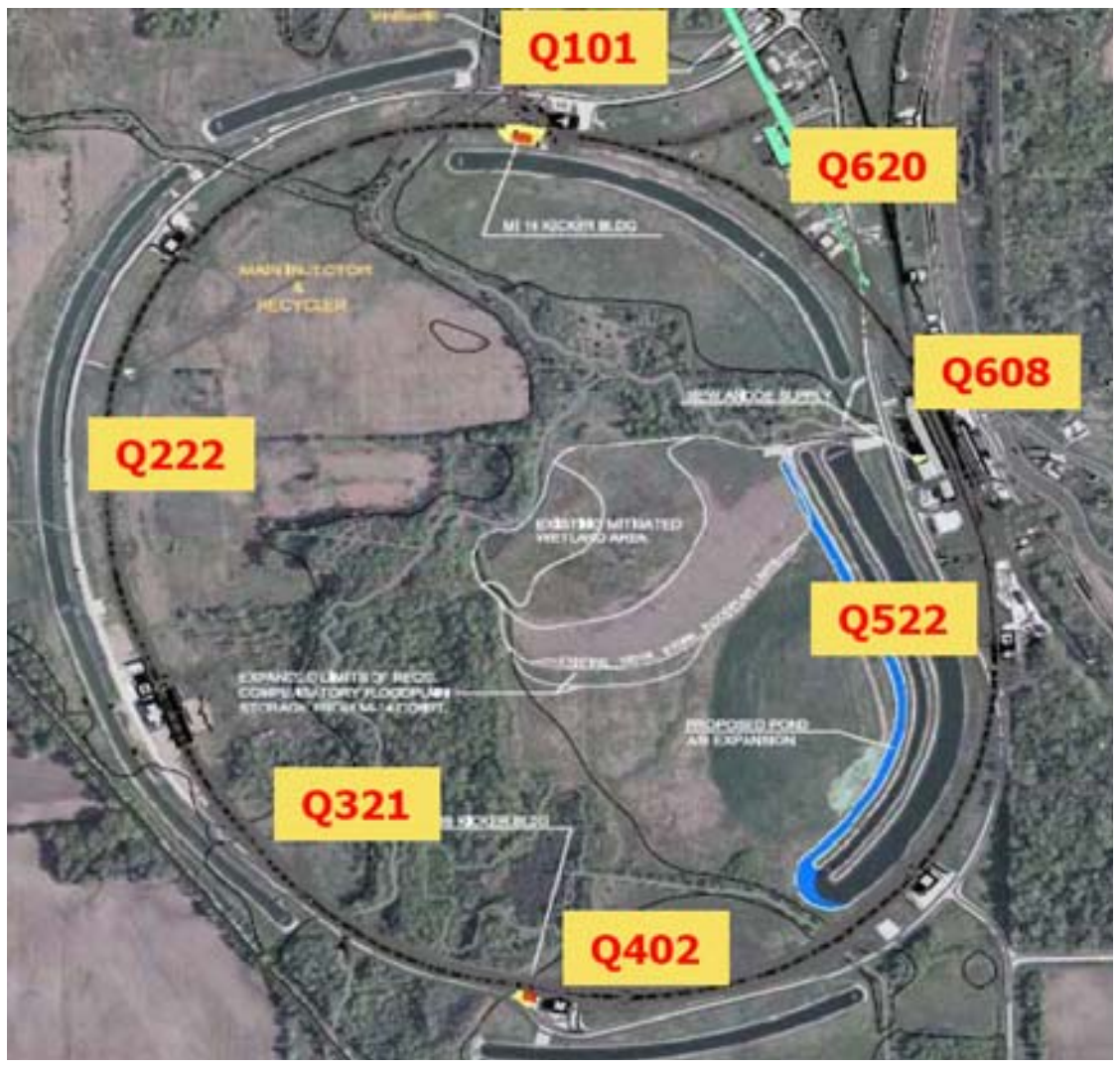

Figure 2: Seven WQBs around the MI ring.

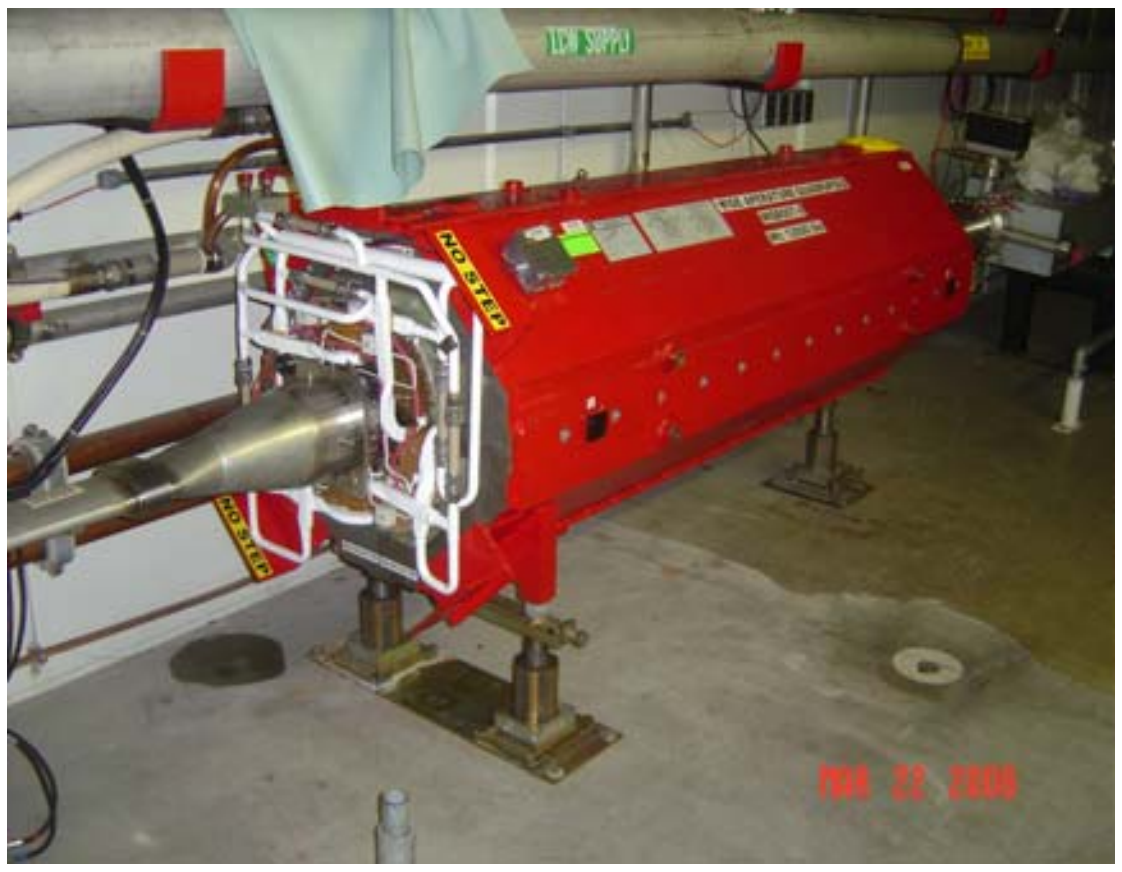

Figure 3: A WQB installed at the Q222 location.

* Work supported by the U.S. Department of Energy under Contract No. DE-AC02-07CH11359.

# chou@fnal.gov

## FIELD ERROR AND CORRECTION

Table 2 lists the main parameters of the WQB.

Table 2: WQB Parameters

| | |
|---|---|
| Aperture | 4.347” |
| Length | 84” |
| Max gradient at 150 GeV/c | 19.6 T/m |
| Good field region | ± 2” |
| Weight | 12,000 lb |
| Main coil | |
| Turns per pole | 7 |
| Peak current at 150 GeV/c | 3540 A |
| RMS current | 2000 A |
| Resistance | 8.1 mΩ |
| Inductance | 3.7 mH |
| Trim coil | |
| Turns per pole | 18 |
| Max current | 28 A |
| Resistance | 0.75 Ω |
| Inductance | 0.03 H |

The WQBs and the MI main quadrupoles are powered by the same buses. If the integrated field error of the WQB was 1%, each WQB would cause a beta-wave of 5.7% and a tune shift of 0.0045, which is not acceptable. There are five horizontal focusing WQBs and two vertical focusing WQBs. In order to minimize the perturbation on the lattice, the allowable WQB field error was set to 0.1%, or 10 units (1 unit = $10^{-4}$).

However, because the WQB has higher saturation (due to larger aperture) and stronger hysteresis (due to steel properties) than the old quads (called the IQB), the measured integrated field error was significantly larger than the specification. It reached +4% at low field and –3% at high field, as shown in Figure 4.

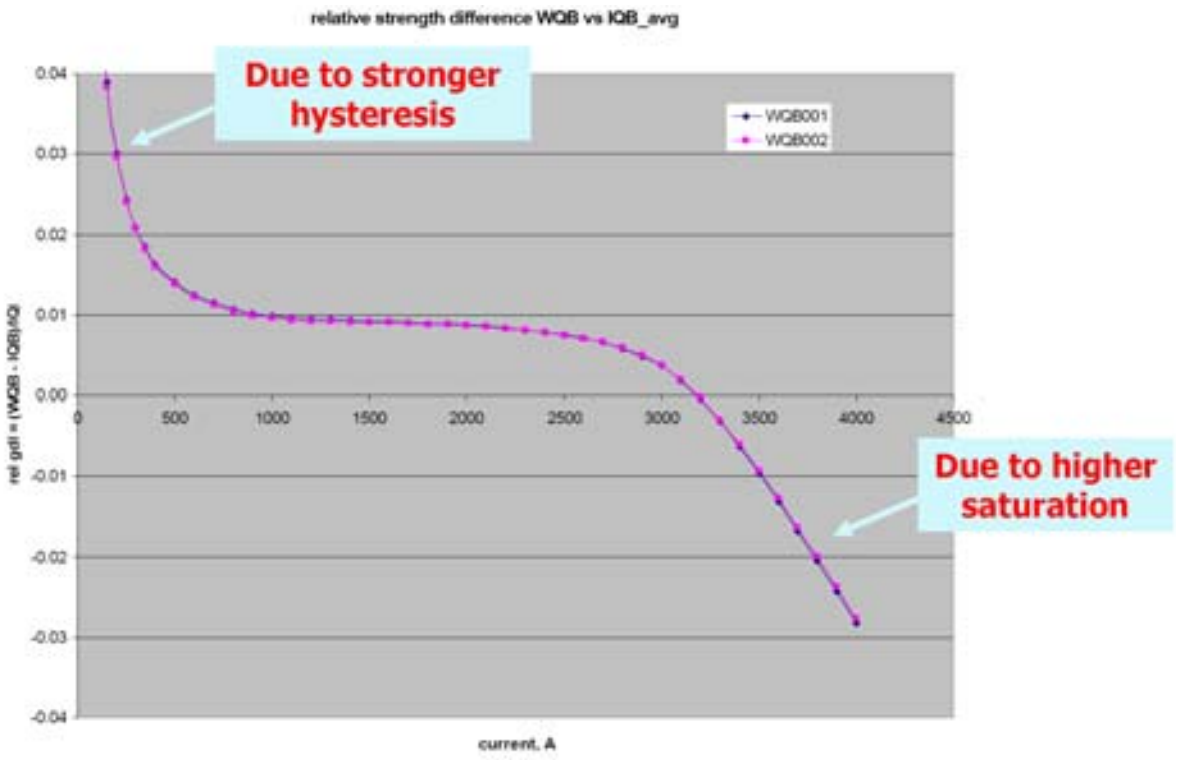

Figure 4: Relative field error of the WQB.

To further complicate the matter, the hysteresis has strong dependence on the reset current (Figure 5) and the transfer function of the trim coil used for field correction has large anomaly at high current. All these were carefully measured and taken into account in designing the required trim current throughout the ramp, which is shown in Figure 6.

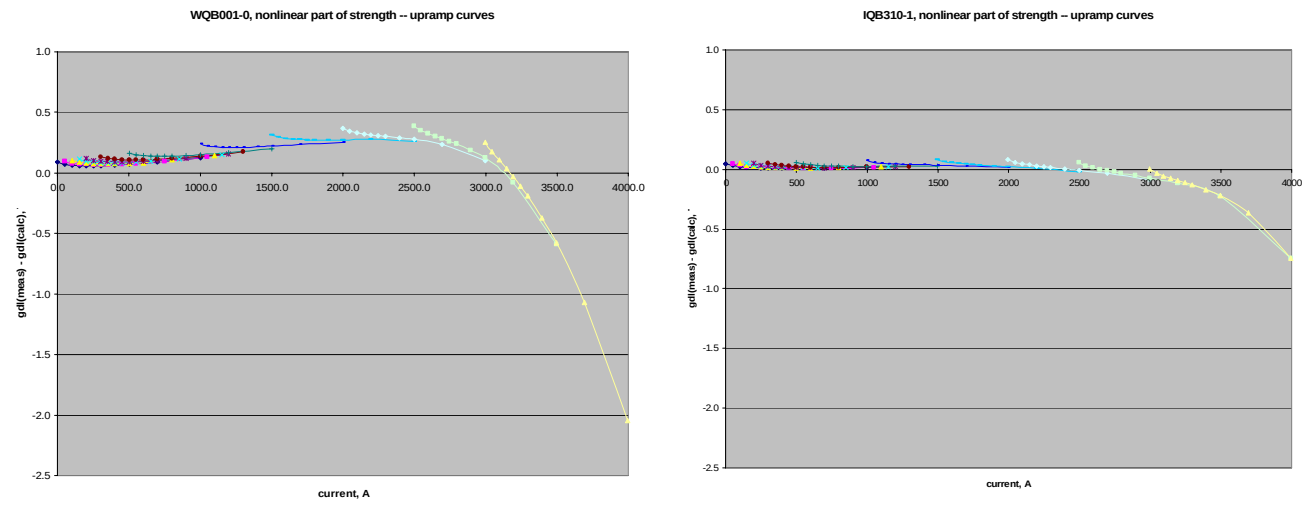

Figure 5: Hysteresis curves at different reset: left – WQB, right – IQB.

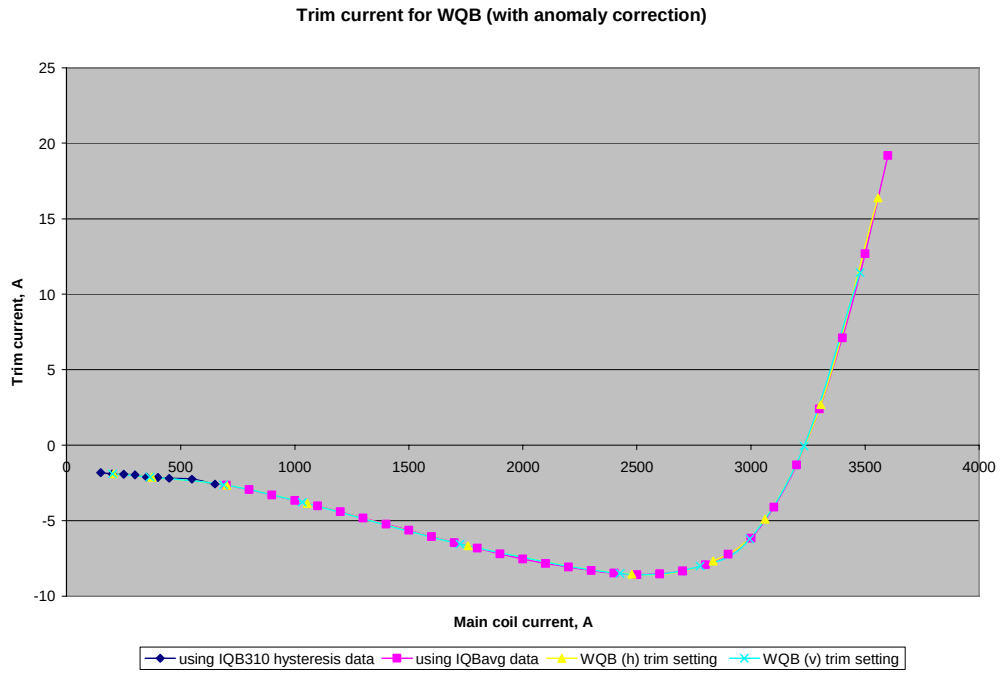

Figure 6: Required trim current during the ramp.

## NEW BPM AND OFFSET TABLE

The new extra wide aperture BPM has an electrode ID of 5.625” and extended angle of each electrode of 60°. It can measure both horizontal and vertical position of a beam at the same location. In order to get an accurate orbit measurement, several offsets have to be corrected:

- Offset between the WQB lamination center and the magnetic field center (~200 μm).
- Survey offset between the mechanical center of the BPM and that of the nearby WQB.
- BPM offset between the electrical and mechanical center of the BPM.
- Electrical offset due to cables, jump boxes and electronics upstairs in the service building.
- Orbit offset which is intentional move of the quadrupoles for accommodating the required large orbit deflection.

All these offsets were measured and incorporated into an offset table in the BPM front-end database.

## ORBIT MEASUREMENT

The simple geometry of the BPM makes it possible to use an analytical formula to compute the beam position from BPM signal [3]:

$$\text{pos (mm)} = A \times R\,\text{(mm)} \times \frac{1-\sqrt{1-x^2}}{x}$$

where $R$ is the BPM electrode radius, $A$ is a constant to be fit to the calibration data, and $x$ is the ratio of the signal difference to the signal sum. There is also another scaling formula suggested by Webber. He used MATHCAD to fit the bench data and obtained a 5[th] order polynomial [4]:

$$pos\,(mm) = 43.513x + 5.432x^3 + 21.071x^5$$

Both formulae were used for data processing and compared with the calibration data. Figure 7 shows an example of the horizontal fit. While the analytical formula is more accurate at small amplitude (< 20 mm), the 5th order polynomial fits better at large amplitude (> 20 mm). Because the beam displacement is big at the WQB locations, it was decided to use the polynomial for data processing.

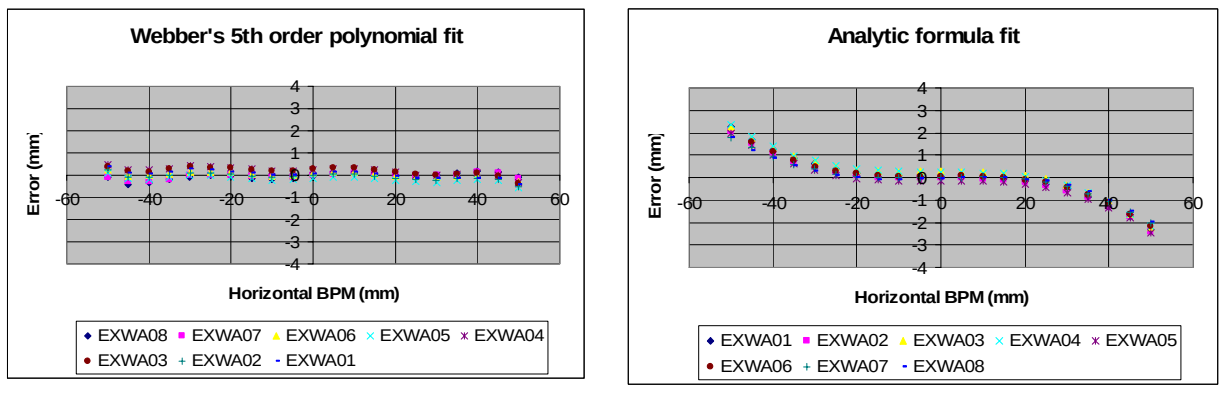

Figure 7: Scaling formula: left – 5th order polynomial, right – analytical formula.

Two types of orbit measurements were carried out to verify the BPM accuracy and estimate the lattice perturbation by WQBs, one using 3-bump method, another 1-bump method. The calculation of closed orbit deviation (COD) of the two methods is well known [2]. The results show a good agreement between the calculation and measurement. As an example, Table 3 gives a comparison of 3-bump COD at 4 WQB locations.

Table 3: Calculation vs. Measurement (mm/0.25A)

| Location | COD (calc.) | COD (meas.) |
|---|---|---|
| Q101 | 1.25 | 1.48 |
| Q222 | 3.21 | 3.44 |
| Q321 | 1.43 | 1.56 |
| Q402 | 3.22 | 3.75 |

## MACHINE APERTURE AND ACCEPTANCE IMPROVEMENT

Figure 8 shows the cross section of the Lambertson magnet, the old beam pipe for the IQB and new pipe for the WQB. The aperture increase is about 10 mm, which was verified by aperture scanning as shown in Figure 9.

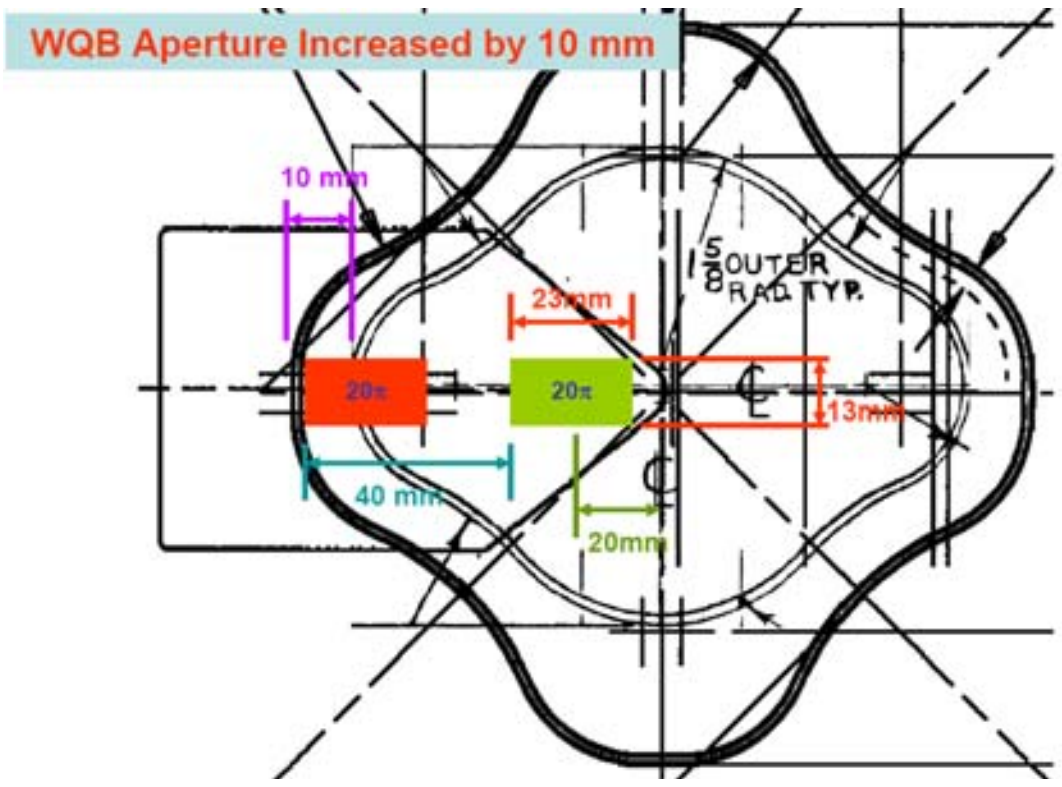

Figure 8: Cross section of Lambertson, old pipe in the IQB and new pipe in the WQB.

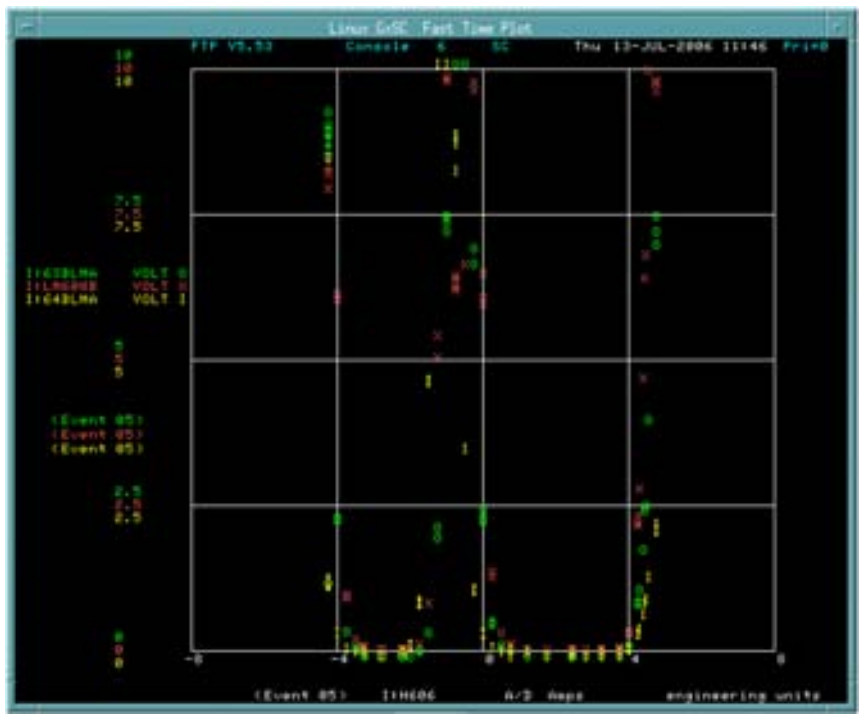

Figure 9: Aperture scanning at Q608.

Figure 10 (top) shows the WQB increases the machine acceptance to 60π mm-mrad, 50% larger than before (Fig. 1). This acceptance is limited by the WQB good field region (2”). If one would move the WQB by 10 mm, the acceptance could be further increased to 80π mm-mrad, as demonstrated in the bottom of Figure 10.

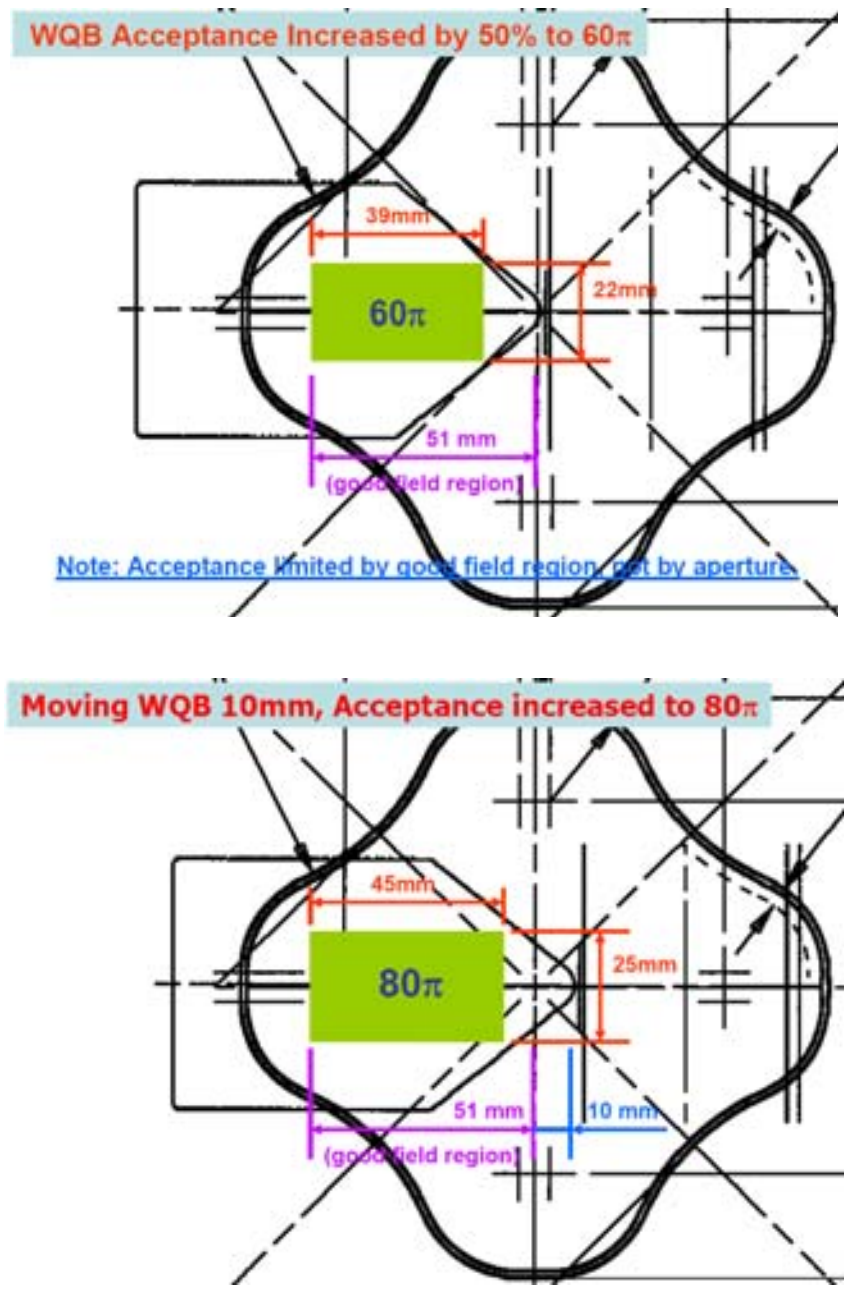

Figure 10: Top – present acceptance, bottom – acceptance with a move of WQB by 10 mm.

## REFERENCES

[1] D. Harding et al., “A Wide Aperture Quadrupole for the Fermilab Main Injector Synchrotron,” this conference.

[2] W. Chou, “Operational Aspects of the MI Large Aperture Quadrupole (WQB),” Fermilab Beams-doc-#2479-v2 (2006), http://beamdocs.fnal.gov/AD-public/DocDB/DocumentDatabase

[3] W. Chou, “Derivation of the S-Curve of BPM Signals,” APS/IN/ACCPHY/89-3, APS Project, Argonne National Laboratory (1989).

[4] R. Webber, “MI Extra Wide Aperture BPM Scaling in Difference-Over-Sum BPM System,” Fermilab Beams-doc-#2479-v2 (2006).